\def\sgm{$\sigma$}
\def\mag{\hbox{$\;.\!\!\!^m$}}
\def\to{\hbox{$\,$--$\,$}}
\def\muspc{\hskip 0.15 em}

\hyphenpenalty=50

\input psfig 
\input aa.cmmb
\pageno=1

%%%%%%%%%%%%%%%%%%%%%%%%%%%%%%%%%%%%%%%%%%%%%%%%%%%%%%%%%%%%%%%%%%%%%%%
\MAINTITLE{Probing the Galaxy 
\newline
I. The galactic structure towards the galactic pole}
\AUTHOR{Y.K. Ng@{1,2,5{\rm (present \ address)}}, 
G. Bertelli@{3,4}, C. Chiosi@3, and A. Bressan@5}
\INSTITUTE{
 @1
 Leiden Observatory, P.O. Box 9513, 2300 RA \ Leiden, the Netherlands
 @2
 IAP, CNRS, 98 bis Boulevard Arago, F-75014 Paris, France
 @3
 Department of Astronomy, Vicolo dell'Osservatorio 5, I-35122 Padua, Italy
 @4
 National Council of Research, CNR -- GNA, Rome, Italy
 @5
 Padova Astronomical Observatory, Vicolo dell'Osservatorio 5, I-35122 Padua, 
 Italy
 @{\null}
 {E-mail:\ {\tt  (bertelli,bressan,chiosi,yuen)\char64astrpd.pd.astro.it}}
}
 
\THESAURUS{04 (3.13.2, 08.08.1, 08.19.1, 10.19.2, 10.19.3)}
\DATE{Received 9 November 1995 / Accepted 20 January 1997}
%%%%%%%%%%%%%%%%%%%%%%%%%%%%%%%%%%%%%%%%%%%%%%%%%%%%%%%%%%%%%%%%%%%%%%%
\ABSTRACT{Observations of (B--V) colour distributions
towards the galactic poles are compared with     
those obtained from synthetic colour-magnitude diagrams
to determine the major constituents in the disc and spheroid.
The disc is described with four stellar sub-populations: 
the {\it young, intermediate, old}\/
and {\it thick disc} populations, which have respectively 
scale heights of 100 pc, 250 pc, 0.5 kpc, and 1.0 kpc.
The spheroid is described with stellar contributions from the {\it bulge}\/
and {\it halo}.
The bulge is not well constrained 
with the data analyzed in this study.
A non-flattened power-law describes the observed distributions
at fainter magnitudes better than a deprojected R$^{1/4}$--law.
Details about the age, metallicity, and normalizations 
are listed in Table~1.
\par
The star counts and the colour distributions from the stars
in the intermediate fields 
towards the galactic anti-centre are well described 
with the stellar populations mentioned above. 
Arguments are given that the actual solar offset is about 
15~pc north from the galactic plane.}
\KEYWORDS{
methods: data analysis -- Stars: general, HR-Diagram, statistics
 -- Galaxy: stellar content, structure }
\maketitle

\MAINTITLE{Probing the Galaxy}
\SUBTITLE{I. The galactic structure towards the galactic pole}
\hyphenation{Paper}
\titlea{Introduction}
Close to the galactic midplane one is most sensitive to stars
with a low or moderate scale height. Information about stellar 
populations from an extended or thick disc can be %tightly 
constrained with observations at high galactic latitudes.
The star counts from the north galactic pole field are well suited 
%for probing the Galaxy. Especially, 
to disentangle and to determine
the scale heights for the various disc populations. 
The apparent scale height for
each spectral type is the cumulative result from different populations
with each their own characteristic
age, metallicity and scale height. 
For the determination
of the scale heights one should start with
the youngest population with the lowest scale height and proceed 
backwards to older populations with larger scale heights.
\par 
Observational data for the north and south galactic pole
(hereafter respectively NGP and SGP)
are available from
several studies: 
Weistrop (1972); Faber et al. (1976);
Chiu (1980); Kron (1980); Koo \& Kron (1982);
Gilmore, Reid \& Hewett (1985, hereafter GRH85);
Murray et al. (1986),
Yoshii, Ishida \& Stobie (1987);
Stobie \& Ishida (1987, hereafter SI87); Reid (1990, hereafter RD90); 
Majewski (1992); Soubiran (1992);
and Reid \& Majewski (1993, hereafter RM93). In this
study
the data for the NGP are taken 
from SI87, RD90, and RM93 and the star counts for
the SGP are taken from GRH85. \par
In the direction of the galactic anti-centre (hereafter GAC) data 
are available from various sources: 
Friel \& Cudworth (1986)
in the direction of ($l,b$)\muspc=\muspc($175\degr,-49\degr$),
Fenkart \& Esin-Yilmaz (1983) and 
Yamagata \& Yoshii (1992, hereafter referred to as YY92)
in the direction of 
\hbox{($l,b$)\muspc=\muspc($200\degr,59\degr$)},
and Ojha et al. (1994, hereafter referred to as OBRM94)
for the field towards \hbox{($l,b$)\muspc=\muspc($167\fdg5,47\fdg4$)}. 
For the analysis we used the data sets from
YY92 and OBRM94.
\par
%A comprehensive study with other galactic models 
%and discussions about the results obtained thus far 
%from various fields can be found in Bahcall \& Soneira (1984),
%Robin \& Cr\'ez\'e (1986), OBRM94 and Robin et al. (1996).
%\par
Star counts from
Selected Areas (Blaauw \& Elvius 1965)
and well studied fields are analyzed with the HRD-GST 
(HR-Diagram Galactic Software Telescope, see Ng 1994 \& 1996 and Ng et al. 1995
for details, the latter is hereafter referred to as Paper~I).
The aim is
to generate synthetic CMDs 
(Colour-Magnitude Diagrams) from first principles,
without putting {\it \`a priori}\/ observed quantities in the model
and to determine in a self-consistent way the 
structural parameters (scale height, flattening of bulge \& halo)
and the evolutionary status (age-metallicity)
of the stellar populations from fields directed towards 
the outer part of
our Galaxy. 
The results are used as input for an overall galactic model
and are then applied to fields in the direction of the SGP and GAC.
In Sect.~2 a description is given of the method used for the analysis.
The results are presented in Sect.~3 and they are discussed in Sect.~4.
\newcolumn

\titlea{Analysis}

\titleb{Method}
The stellar population synthesis technique used for the HRD-GST
is a powerful tool for the understanding
of the complex distribution of stars in CMDs.
The different stellar evolutionary phases 
are linked to each other through libraries
with stellar evolutionary tracks.
An interpolation is made between sets of
evolutionary tracks with different metallicities
to obtain a smooth metallicity coverage.
With assumptions about the age, metallicity and the shape of the IMF
(initial mass function), one obtains information about the star 
formation history (Bertelli et al. 1992) and the (synthetic) 
luminosity function. 
\par
The MS (main sequence) up to the AGB (asymptotic giant branch) stars 
from a synthetic population cover a specified
age and metallicity range. The mass spectrum 
is specified with a power-law IMF and
the number of stars with a specific age is determined by
the SFR (star formation rate).
The stars in a particular population all have the same scale
height, because they are formed at the same period. 
High mass stars are only present in (very)
young populations, while low mass stars are present in all
populations under consideration. 
\par
The HRD-GST has thus far been used for the interpretation of the star counts 
from fields towards the galactic centre
(Paper~I, Bertelli et al. 1995 \& 1996,
the first reference is hereafter referred to as Paper~II; Ng et al. 1996a,
hereafter referred to as Paper~III).
In Paper~I and Ng (1994) it is demonstrated that the age, the metallicity,
the scale height of the disc stellar populations 
and the spatial distribution of the bulge/halo populations 
(flattening parameter) are the most sensitive parameters.
They do not depend critically on the exact choice 
of the remaining input parameters for which we adopt
reasonable values; see Table 1 for the scale length,
IMF and SFR.
The latter parameters have not been changed throughout the analysis.
%The analysis of those parameters is beyond the scope of this paper
%and will be subject for a future paper.
In this way, the study is simplified and one can focus on the exploration
of a limited, but fundamental fraction, of the HRD-GST parameter space. 
The parameters are determined through an iterative approach:
first the most sensitive 
parameter then the next sensitive and so forth.
\par
The combined data sets from the selected NGP fields 
span a range of 12 magnitudes
in V. We determine the scale heights and age-metallicity
for the various stellar populations along the line of sight
and apply them to fields towards the SGP and 
to intermediate latitude fields towards the GAC.
It is not possible to obtain stronger constraints for
the scale heights and age-metallicity from the latter fields, 
because they cover a smaller magnitude range. 
Those fields provide
a consistency test and an independent verification of the input
parameters from the HRD-GST for an overall galactic model.
In all fields the same local normalization is applied (see Sect.~2.3),
as determined from the NGP star counts (see Sect.~3.1).
Any difference between model and observations is either 
an imperfection of the model or 
is due to a local difference in the galactic structure.

\titleb{Limits}
We adopted 
in the MC (Monte-Carlo) simulations for all fields 
the following limiting magnitudes:
\hbox{B$_{lim}$\muspc=\muspc23\mag5}, and 
\hbox{V$_{lim}$\muspc=\muspc22\mag0}.
Differences in the definition of 
zero-points of the photometric passbands 
and their transformations result in small 
colour shifts between the simulated and observed data. 
Shifts are also induced by differences in age, metallicity
or extinction and are considered
when a general shift due to a zero-point difference did not remove
the observed discrepancy.
\par
The parameters for the HRD-GST determined in this paper 
are not unique. This 
makes any test as valid as the most simple test: 
the MC-simulations goes through all the data points.
It has been assumed 
that young disc populations ought to have higher metallicities
and lower scale heights than older ones. 
%This is equivalent with the 
%assumption, that the disc formed through a gravitational collapse, where 
%the ejecta of the evolved, high mass stars gradually enriched 
%its environment.
\par
The present mass limit is 0.6 M/M$_\odot$ and 
work is in progress to extend it to lower masses. 
(Girardi et al. 1996a,b). 
The absence of synthetic main sequence stars with even lower masses 
%(B--V\muspc$>$\muspc0\mag95 for metal poor stars and 
%B--V\muspc$>$\muspc1\mag35 for solar metallicity,
%see respectively Figs.~2.4 \& 2.1) 
is not a major disadvantage.
The missing stars do not contribute to the evolved stars, because
they remain on the main sequence for at least a Hubble time.
At bright magnitudes, say \hbox{V\muspc$<$\muspc14$^m$}, 
the contribution of low mass stars is negligible.
At faint magnitudes, say V\muspc$>$\muspc18$^m$,
the colour distribution splits up around B--V\muspc$\simeq$\muspc1\mag0
in two parts: a blue and a red distribution (see Fig.~1.3).
We have tied our calibration to the blue section, where 
the low mass stars do not contribute. The presence
of low mass stars is not necessary for the determination
of the scale heights and the age-metallicities of the disc populations. 
For the latter a calibration to the blue section is essential.

\titleb{Normalization}
The total mass of the stars within
0.5~kpc along the line of sight is determined for each population.  
The various populations are then scaled relative
to the intermediate disc population (see Table 1). A normalization
based on the local mass density in the solar neighbourhood for each
population is used here, because it is independent of the assumed passband
and detection limit of the simulations. It can eventually be tested 
against the mass density in the solar neighbourhood.
The contribution for each population is obtained by matching 
through trial and error the observed and the simulated distribution.
The following local normalization
is inferred for the various stellar populations:
[halo/disc]$_{local}$\muspc=\muspc1/250, 
[bulge/disc]$_{local}$\muspc=\muspc1/2850, 
[thick disc/disc]$_{local}$\muspc=\muspc4/75, 
[old disc/disc]$_{local}$\muspc=\muspc7/18, and
[young disc/disc]$_{local}$\muspc=\muspc4.6 .
This normalization has also been applied in the other fields
shown in Sect.~3.
The normalization above is governed by the local mass density 
and is not the same as the normalization to the local stellar density 
from other star counts models.
\par
\begtabfull
\tabcap{1}{The stellar populations in the
HRD-GST determined from the analysis of the star counts
towards the North Galactic Pole. 
It is emphasized that the shapes for the IMF and SFR 
have been assumed and have not been determined from the 
star counts data. 
For all populations a power-law IMF has been adopted with an index 
$\alpha\,$=\muspc2.35. For the disc populations a scale length 
of 4.5~kpc is assumed. Better constraints on this
value cannot be obtained from this study.
}
\vbox{\petit\baselineskip 9pt%
\centerline{\vbox{%
\halign{#\hfil&\quad#\hfil \cr
\noalign{\hrule\vskip0.05truecm\hrule\smallskip}
Solar Position:&\cr
\noalign{\medskip}
\qquad Distance from the plane &45$\pm$5 pc north$^\dagger$\cr
&\cr
Stellar Populations:&\cr
\noalign{\smallskip}
\quad{\it Halo}&\cr
\qquad Density law           & Power-law with index n\muspc=\muspc3.0 \cr
\qquad Axial ratio $q$       & 1.0\cr
\qquad Z                     & 0.0004\to0.003\cr
\qquad t                     & 16\to10 Gyr\cr
\qquad SFR                   & Constant\cr
\noalign{\smallskip}
\quad{\it Bulge}$^{\dagger\dagger}$&\cr
\qquad Density law           & Power-law with index n\muspc=\muspc3.0 \cr
\qquad Axial ratio $q$       & 1.0\cr
\qquad Z                     & 0.005\to0.080\cr
\qquad t                     & 15\to13 Gyr\cr
\qquad SFR                   & Exponentially decreasing\cr
\qquad                       & Characteristic time-scale 2.0 Gyr\cr
\noalign{\medskip}
\quad{\it Thick Disc}$^*$&\cr
\qquad Scale height          & 1.0\muspc$\pm$\muspc0.1 kpc\cr
\qquad Z                     & 0.0004\to0.003\cr
\qquad t                     & 16\to10 Gyr\cr
\qquad SFR                   & Constant\cr
\noalign{\smallskip}
\quad{\it `Old' Disc}&\cr
\qquad Scale height          & 0.50\muspc$\pm$\muspc0.05 kpc\cr
\qquad Z                     & 0.003\to0.008\cr
\qquad t                     & 10\to7.0 Gyr\cr
\qquad SFR                   & Exponentially decreasing\cr
\qquad                       & Characteristic time-scale 3.0 Gyr\cr
\noalign{\smallskip}
\quad{\it Intermediate Disc}&\cr
\qquad Scale height          & 0.25\muspc$\pm$\muspc0.02 kpc\cr
\qquad Z                     & 0.008\to0.015\cr
\qquad t                     & 7.0\to4.5 Gyr\cr
\qquad SFR                   & Exponentially decreasing\cr
\qquad                       & Characteristic time-scale 2.5 Gyr\cr
\noalign{\smallskip}
\quad{\it Young Disc}&\cr
\qquad Scale height          & 0.10\muspc$\pm$\muspc0.01 kpc\cr
\qquad Z                     & 0.015\to0.020\cr
\qquad t                     & 5.0\to1.0 Gyr\cr
\qquad SFR                   & Exponentially increasing\cr
\qquad                       & Characteristic time-scale 4.0 Gyr\cr
&\cr
Local Normalization$^{**}$:& |R|$\,<\,$0.5 kpc\cr
\noalign{\smallskip}
\qquad halo/disc             & 1/250 \cr
\qquad bulge/disc            & 1/2850 \cr
\qquad thick disc/disc       & 4/75 \cr
\qquad old disc/disc         & 7/18 \cr
\qquad young disc/disc       & 4.6 \cr
\noalign{\medskip\hrule\vskip0.05truecm\hrule\medskip}
}}}
\noindent
\hbox to 0.5cm{$^\dagger$\hfill} The value for the solar offset
determined is too high; 
15~pc is a more likely value, see Sect. 4.3 for details.\newline
\hbox to 0.5cm{$^{\dagger\dagger}$\hfill}
The parameters have been adopted from Paper II, but its presence 
is not firmly established from the north galactic pole field. 
\newline 
\hbox to 0.5cm{$^*$\hfill} The age of this population is 
probably about 1\to3 Gyr 
too old, see Sect. 4.4.2 for details.\newline
\hbox to 0.5cm{$^{**}$\hfill} This normalization is made with respect to 
the intermediate disc. We normalize to a local mass density, 
see Sect. 2.3 for details.
}
\endtab

\titlea{Results}
Table 1 lists a description of the stellar populations, which are 
determined with the HRD-GST from mainly 
the NGP star counts.
A total of six different synthetic stellar populations
(halo, `bulge', thick disc, old disc, intermediate disc, and young disc)
are found from the analysis of the NGP data set.
The bulge population might not be essential. 
\par

\titleb{The north galactic pole}
\centerline{NGP: $l$\muspc=\muspc81$\fdg$0, 
$b$\muspc=\muspc+87$\fdg$0 
\ \& \ 
SA 57: $l$\muspc=\muspc65$\fdg$5, $b$\muspc=\muspc+85$\fdg$5}
\medskip
The surveys from SI87, RD90, and RM93 cover an area of respectively 
21.46 degree$^2$, 28.3 degree$^2$ and 0.29 degree$^2$.
\hbox{Majewski~(1993)} pointed out that a gradient in the magnitude scale 
as large as 0\mag25 is present in the Chiu (1980)
magnitudes. This gradient is probably caused by
systematic errors in the old iris photometry.
Because RD90 based his photometric calibration on the Chiu (1980) 
magnitudes, systematic errors can be present in the photometry. \par
In the analysis the following offsets in (B--V) are adopted
in the MC simulations:
+0\mag05 and \to0\mag05 (+ refers to a blue shift of the
MC distribution) 
for respectively SI87 and RM93. 
In the data set from RD90 a small gradient is present, resulting
in a (B--V) shift of +0\mag15 for 
\hbox{V\muspc=\muspc13$^m$\to14$^m$} and +0\mag05
for 
\hbox{V\muspc=\muspc18$^m$\to19$^m$}. An offset of +0\mag10 
is adopted in Fig.~1.2.
\par
For the NGP we adopted 
E(B--V)\muspc=\muspc0\mag0.
For the simulations of the 
photometric errors the following values are used
over the whole magnitude range of interest:
\hbox{\sgm(B)\muspc=\muspc0\mag02} and 
\hbox{\sgm(V)\muspc=\muspc0\mag02} 
for the RM93 star counts,
\hbox{\sgm(B)\muspc=\muspc0\mag05} and 
\hbox{\sgm(V)\muspc=\muspc0\mag05} 
for the SI87 star counts, and
\hbox{\sgm(B)\muspc=\muspc0\mag10} and 
\hbox{\sgm(V)\muspc=\muspc0\mag10} 
for the RD90 star counts.
Comparable errors need to be adopted in the MC-simulations,
otherwise a comparison with the observations is meaningless. 
\par
In Figs. 1.1\to1.3 
the colour distributions from 
respectively SI87, RD90 and RM93
are compared with the MC simulations.
In figure 1.1a it is noted that the 
MC simulation over-predicts considerably the number of evolved stars 
around \hbox{B--V\muspc=\muspc1\mag2}. 
Inspection of the stars brighter than V\muspc=\muspc12\mag5 in SI87's CMD 
(their Fig.~7)
shows, that the stellar locus becomes gradually 
bluer towards brighter magnitudes. 
This could be due to the presence of even younger stars,
but also due to systematic errors 
in the calibration of bright stars. The latter moves
the bright evolved stars into the bluer stellar locus. 
\par
Figure 2.0 shows the combined CMD of the six stellar 
populations considered. 
Figures 2.1, 2.2, 2.3, 2.4, 2.5, and 2.6
display the individual CMDs,
respectively
young disc, 
intermediate disc, 
old disc, 
thick disc, 
bulge, and
halo.
Table~1 gives a description for each population.
Figure 2.7 shows the CMD for the halo contribution when 
the \hbox{$R^{1/4}$--law} is used.
%The CMDs show for each population the distribution of main sequence (dots),
%core H-exhausted or red giant branch (dashes), horizontal branch
%(open circles), and asymptotic giant branch (filled squares) stars.
%The object densities in the synthetic CMDs are taken relatively to the
%number of young disc for which 1000 synthetic stars are plotted 
%in Fig.~2.1. 
The simulations in Figs.~2.0\to2.7 represent 
an area of about 5.0 degree$^2$.
In Figs. 2.1\to2.4 the sharp cutoff
of the MS (main sequence) due to the lower mass limit of the evolutionary
tracks in the HRD-GST is clearly visible. Note that this 
cutoff becomes redder, when going from an old and metal-poor 
population (Fig.~2.4) to younger metal-richer populations (Figs.~2.3\to2.1).
From Fig.~2.4 it is seen that the thick disc contribution
becomes negligible for 
\hbox{B--V\muspc$\simeq$\muspc0\mag5} and 
V\muspc$>$\muspc18$^m$. At this colour and magnitude 
the stellar contribution can only be due to halo stars.
A point of attention is that the current description of the 
halo population does not provide blue horizontal branch stars 
(see Fig.~2.6). 
These stars can be generated by lowering the metallicity 
or adopting an older age. The latter shifts the main sequence 
and the core H-exhausted stars to redder colours and increases the 
discrepancy even more. Therefore, an older age for the same metallicity range 
can be ruled out and the lower metallicity limit of the halo
population should most likely be decreased. In this case a slightly
older age is allowed.
\par
\advance\pageno by 2
Figure 3.1c shows from the MC simulations
the resulting bolometric luminosity function
for stars in the spheroid (Fig.~3.1a) and 
in the disc (Fig.~3.1b) component.\par
\begfigps 9.5cm
\psfig{file=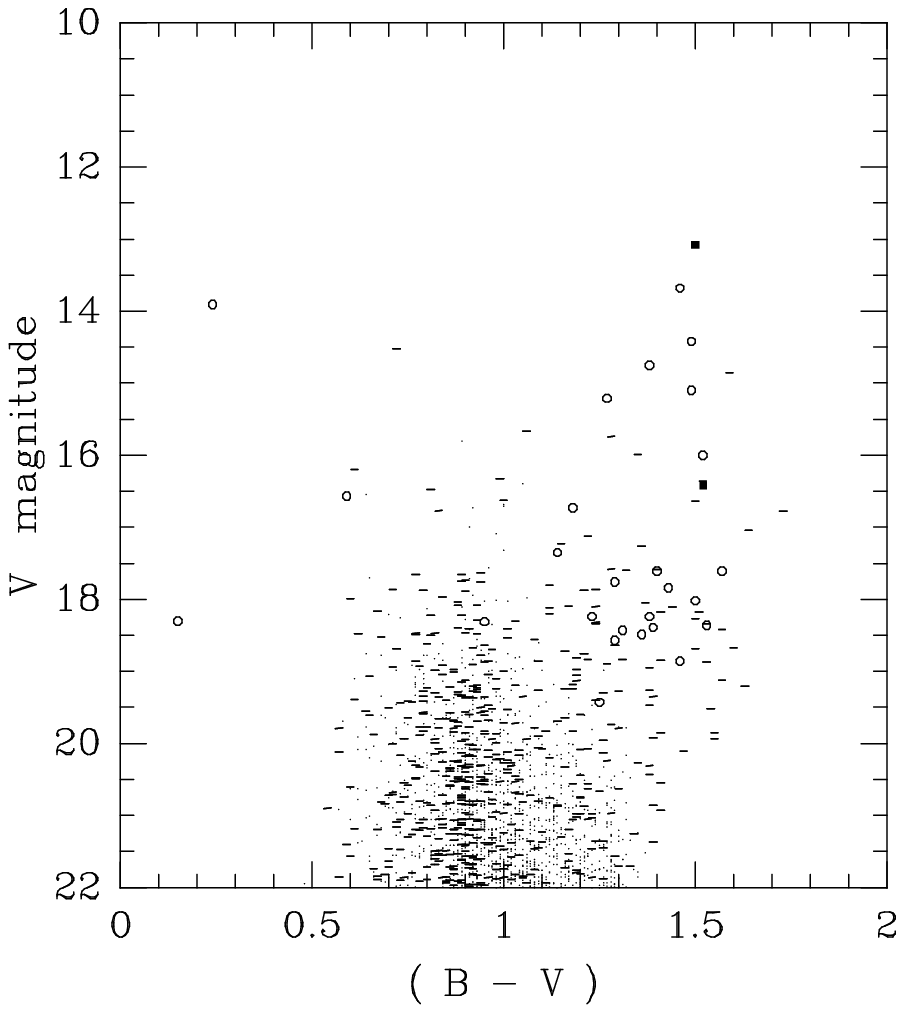,height=9.5cm,width=8.6cm}
\figure{2.5}{Simulated (V,B--V) CMD for the bulge
population in the direction of the NGP,
see figure 2.0 for the symbols and table 1 for details
%\hfill\break
}
\endfig
\begfigps 9.5cm
\psfig{file=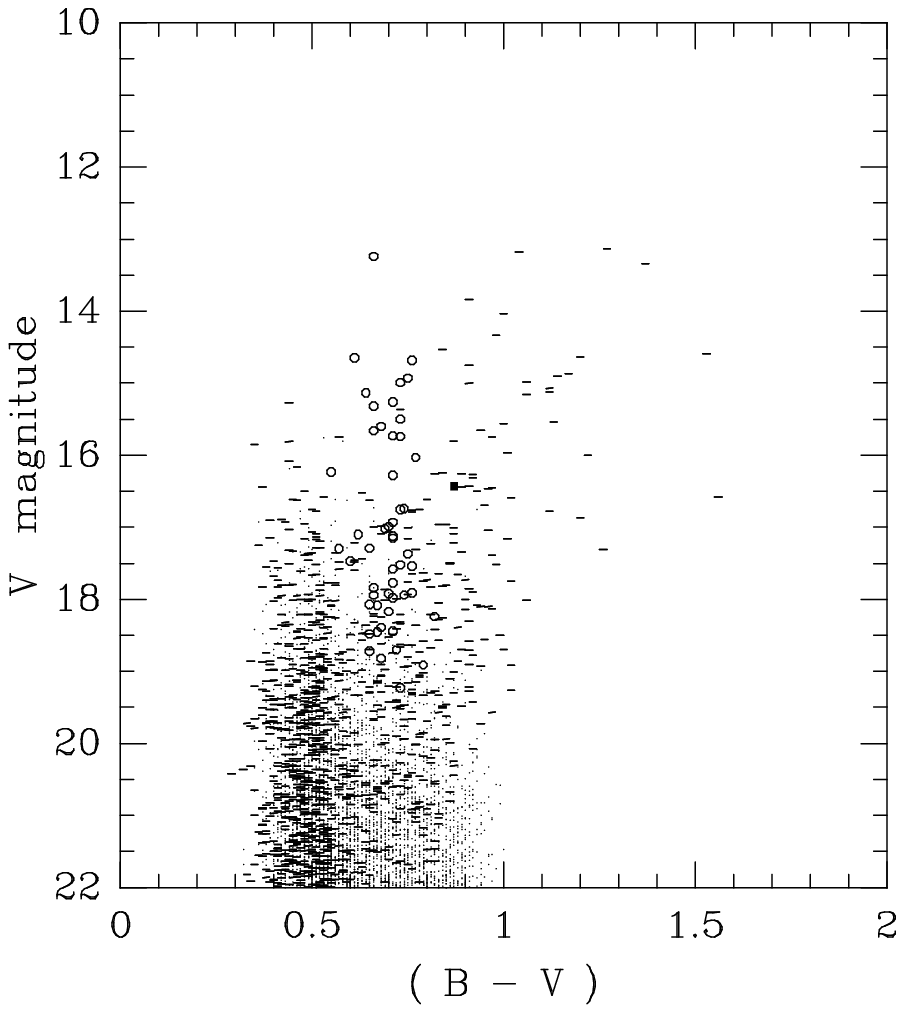,height=9.5cm,width=8.6cm}
\figure{2.6}{Simulated (V,B--V) CMD with a power-law for the halo 
population in the direction of the NGP,
see figure 2.0 for the symbols and table 1 for details}
\endfig
\begfigps 9.5cm
\psfig{file=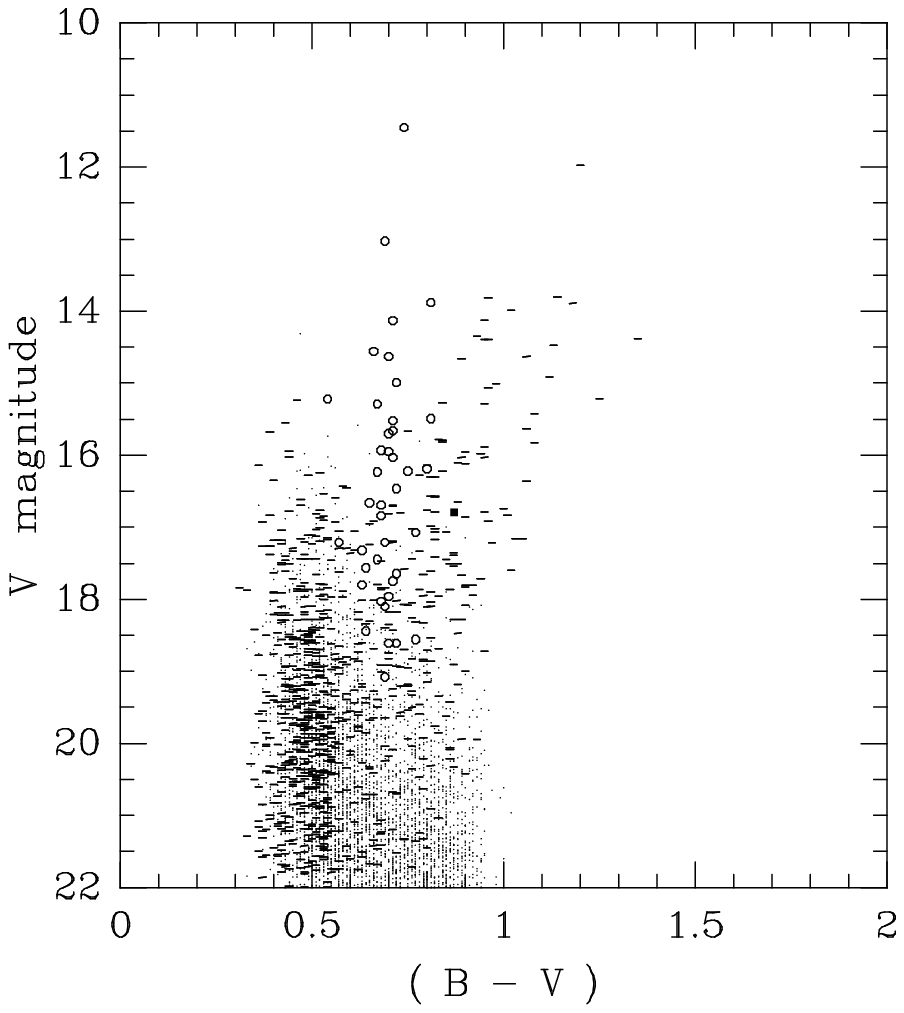,height=9.5cm,width=8.6cm}
\figure{2.7}{Simulated (V,B--V) CMD with the \hbox{$R^{1/4}$--law} 
for the halo population in the direction of the NGP,
see figure 2.0 for the symbols and table 1 for details
\vfill\null
}
\endfig
Figure 3.2c shows the resulting star counts from the MC simulations
in the spheroid (Fig.~3.2a) and in the disc (Fig.~3.2b), together with
the observed star counts from SI87 (filled dots) and RM93 (open dots).
From this figure it is evident that up to 
V\muspc=\muspc18\mag0 the majority of the observed stars
are disc stars. The stars in the spheroid become important 
for \hbox{V\muspc$>$\muspc18\mag0}. In Figs. 1.2d, 1.2e,
1.3a, and 1.3b it can be seen 
that the HRD-GST model over-predicts the number of observed stars, while
low mass main sequence stars with \hbox{B--V\muspc$>$\muspc1\mag4} are lacking.
In the star counts (Fig.~3.2b) one notices that these differences 
cancel each other out.
One cannot rely solely on the 
goodness of star counts fits. An inspection is required
of the colour distributions and/or the synthetic CMD. 
\par
\begfigps 9.4cm
\psfig{file=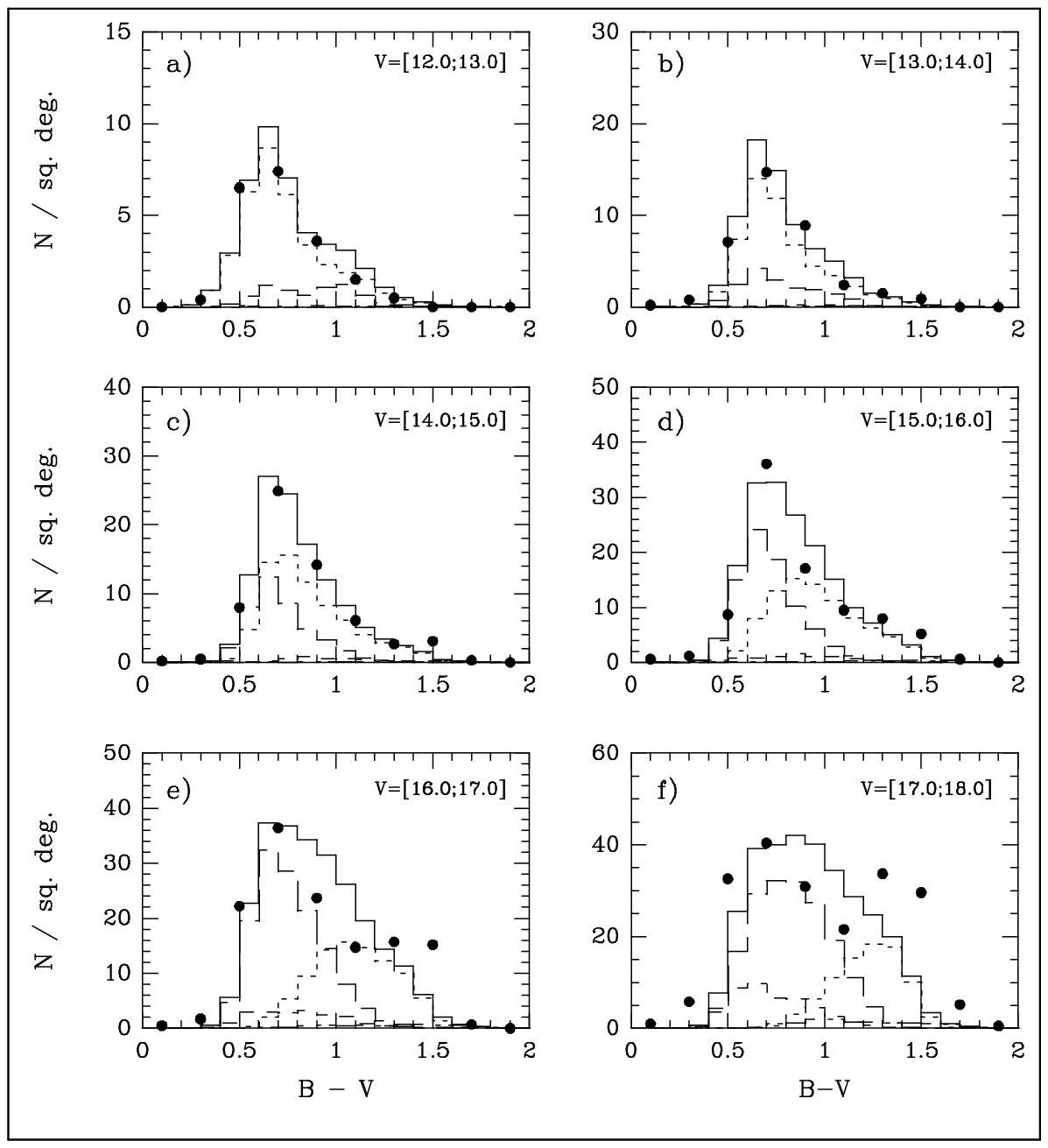,height=9.4cm,width=8.6cm}
\figure{4.1}{Colour distribution from Gilmore,
Reid \& Hewett (1985, filled dots)
in the direction of the SGP; see text and Fig.~1.1 for
additional details}
\endfig
Figure 3.3c gives the total number of simulated stars detected
in the line of sight through the spheroid (Fig.~3.3a) and 
the disc (Fig.~3.3b). This and the next figure suggest,
that there are some nearby (\hbox{d\muspc$<$\muspc250~pc}) 
stars missing in the disc of the HRD-GST model.
One would expect in the local solar neighbourhood 
a gradually increasing or flat curve, but not a 
decreasing one.
%It is not clear, if these missing stars are from the young disc population
%or that they are from an even younger population,
%not yet constrained by or detected in the data analyzed thus far.
%The sampled volume is small at close distance and not
%a large number of stars are detected herein. The majority
%of the stars present have masses \hbox{M/M$_\odot$\muspc$<$\muspc0.6},
%which are currently lacking in the HRD-GST library of evolutionary tracks.
The discrepancy originates
from stars with masses \hbox{M/M$_\odot$\muspc$<$\muspc0.6}.
Notice that there is no dip at the disc/spheroid transition.
%,indicating gravitational equilibrium in consecutive layers. 
\par
\advance\pageno by 1
Figure 3.4c gives the total mass, including 
those below the detection limit, from stars
with 
M/M$_\odot\!\!>\,$0.6
in the line of sight in the spheroid (Fig.~3.4a) and 
in the disc (Fig.~3.4b).
\par
\begfigps 9.4cm
\psfig{file=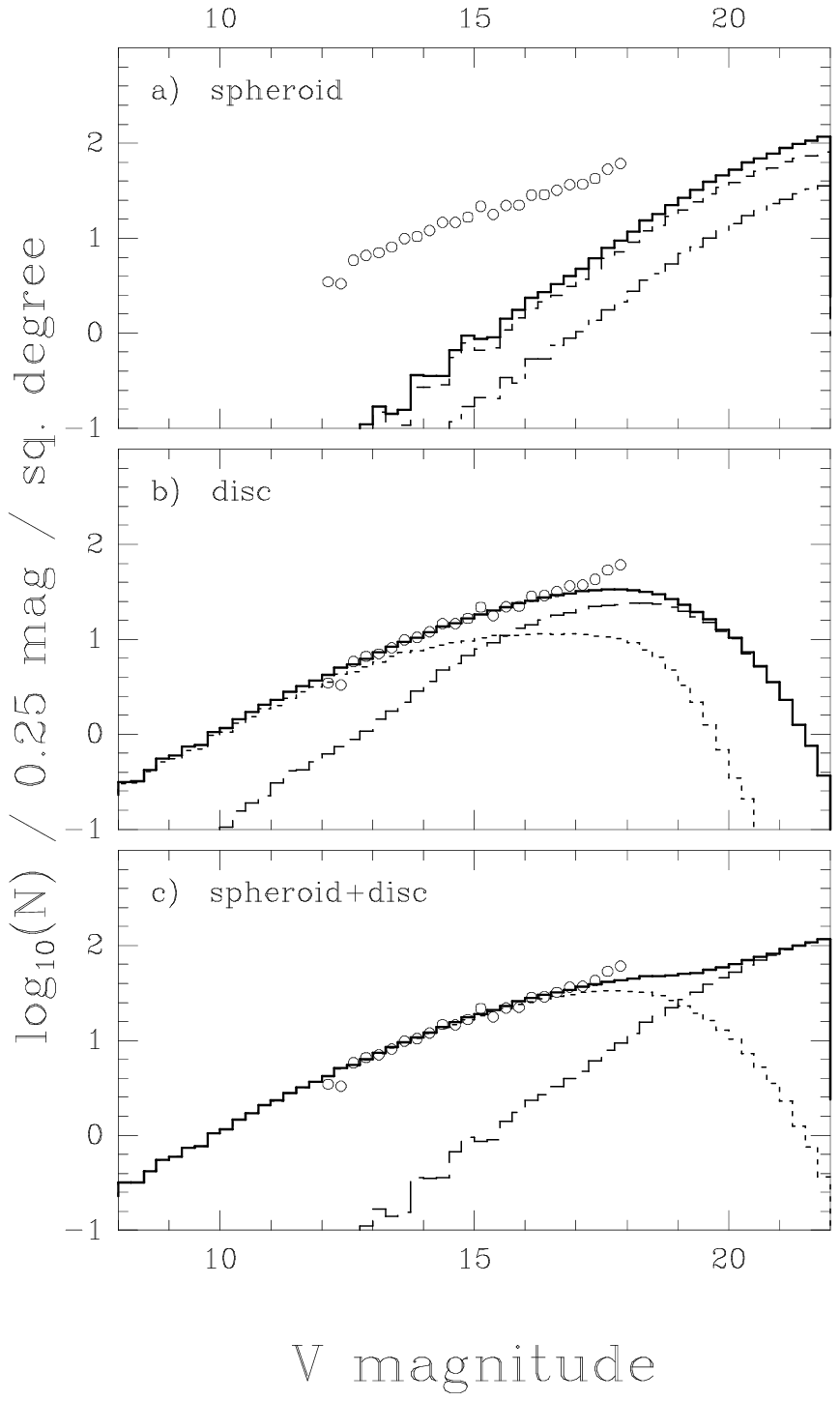,height=9.4cm,width=8.6cm}
\figure{4.2}{Star counts
from the MC simulations shown in Fig.~4.1.
Open dots are the star counts from GRH85,
see captions of the Figs. 1.1 \& 3.1 for
additional details
%\vskip 3.5mm
}
\endfig

\titleb{The south galactic pole}
For the SGP we used the GRH85 star 
counts, covering an area of 18.2 degree$^2$. 
For the MC simulations we found after trial and error:
\hbox{E(B--V)\muspc=\muspc0\mag00}, 
\hbox{\sgm(B)\muspc=\muspc0\mag05} and 
\sgm(V)\muspc=\muspc0\mag05. 
There is no apparent offset present between the 
observed and simulated (B--V) colour distributions.
But it is possible, that the reddening and the offset 
cancel each other out.
\par
Figure 4.1 shows the comparison of the SGP data with the HRD-GST
predictions. 
Figure 4.2c shows the resulting star counts from the MC simulations
in the spheroid (Fig. 4.2a) and in the disc (Fig. 4.2b), together 
with the observations. The GRH85, YY92 and OBRM94  star counts are mainly
due to disc stars. 
One has to be cautious, because 
it is not possible to determine the 
flattening of the spheroid from these data sets.
\par

\titleb{Intermediate latitude fields}
\medskip
\line{3.3a.\hfill SA 54: $l$\muspc=\muspc200$\fdg$1, 
$b$\muspc=\muspc+58$\fdg$8\hfill\qquad}
\smallskip\noindent
The source of the data for SA 54 is YY92.
Their study covered an area of 16 degree$^2$.
For the MC simulations we adopted:
E(B--V)\muspc=\muspc0\mag08, \sgm(B)\muspc=\muspc0\mag10 and 
\sgm(V)\muspc=\muspc0\mag10.
\hfill\break
Figure 5.1 shows the comparison of observational data from 
YY92 with the HRD-GST. 
In Figs. \muspc5.1a\muspc \& \muspc5.1b\muspc 
the colour distributions predicted by
the HRD-GST are shifted respectively 0\mag20 and 0\mag05 blueward
in Figs. \muspc5.1a\muspc and \muspc5.1b\muspc.
The origin of this shift is probably caused by a calibration error,
due to saturation effects of stars brighter than 
\hbox{V\muspc=\muspc12$^m$}.
Additional observations in independent directions are 
required to confirm this. Because it might 
on the other hand, not be due to systematic photometric errors. 
Exactly the same shift is present in the colour distribution
from OBRM94, who used comparable photographic
material in their study. 
Figure 5.2c shows the resulting star counts from the MC simulations
in the spheroid (Fig.~5.2a) and in the disc (Fig.~5.2b), together 
with the observations.

\medskip
\line{3.3b.\hfill $l$\muspc=\muspc167$\fdg$5, 
$b$\muspc=\muspc+47$\fdg$4\hfill\qquad}
\smallskip\noindent
The source of the data for this field is OBRM94.
Their study covered an area of 18.8 degree$^2$.
For the MC simulations the values for the reddening and 
photometric errors are the same as those adopted for YY92.
\hfill\break
Figure 6.1 shows the comparison of the
OBRM94 data with the HRD-GST predictions. 
In Figs. 6.1a \& 6.1b the colour distributions predicted by
the HRD-GST are shifted respectively 0\mag20 and 0\mag05 blueward.
Figure 6\to2c shows the resulting star counts from the MC simulations
in the spheroid (Fig.~6.2a) and in the disc (Fig.~6.2b), together 
with the observations.
\par
\advance\pageno by 1
A comparison of Fig.~6.2 with Fig.~10 from OBRM94
shows that the HRD-GST model basically covers the whole
magnitude range of interest, while the Besan\c con model
over-predicts 
the number of stars 
in their Fig.~5 at \hbox{V\muspc$<$\muspc13$^m$}.
The origin of this is not clear, because the (B--V)
distributions for \hbox{V\muspc$<$\muspc13$^m$}
are not displayed by OBRM94. Possibly the contributions
from the young and intermediate disc populations have been 
slightly over-estimated
in the Besan\c con model, because relative higher weights
are given to the \hbox{V\muspc$>$\muspc13$^m$} magnitude bins.
A comparison between OBRM94's Fig.~11a
with our Fig.~6.1d shows that the HRD-GST under-predicts
the colour distribution slightly, while the Besan\c con model
is in good agreement. 
The discrepancy with the HRD-GST predictions is 
caused by the relative normalization between the young and intermediate
disc.
\par 

\titlea{Discussion}

\titleb{Age and metallicity}
The stellar populations from the disc and spheroid have been 
successfully used in star counts studies in different directions 
in our Galaxy. The same local normalization is used for all fields.
No deviations have been found, which are directly related with metallicity 
differences in one of the stellar populations. 
\par
Figure~7 shows 
the age-metallicity relation for the disc populations
listed in Table~1.
This relation decreases in metallicity to old age. 
Freeman (1992) demonstrated
with the Edvardsson et al. (1993) sample of F-stars
that there is no apparent evidence for a gradient
in the age-metallicity relation. There is mainly a 
large scatter in metallicity at a particular age. 
However Edvardsson et al. (1993) used the Vandenberg (1985) isochrones 
to determine the ages. Those isochrones were computed with 
opacities derived from the LAOL (Huebner et al. 1977) element mixes.
Significant changes in the ages of intermediate to solar metallicity
stars are expected, when they are derived from isochrones computed with the 
radiative opacities derived from OPAL (Rodgers \& Iglesias 1992, Iglesias
et al. 1992). At \hbox{Z\muspc=\muspc0.001} there is virtual no age
difference between the isochrones computed with LAOL or OPAL based
opacities (Bertelli et al. 1994). Towards higher metallicities 
the ages will be progressively younger. 
The flat age-metallicity relation might be an artifact
of LAOL based isochrones. At present the situation is not clear 
and a revision of the ages from the Edvardsson et al. data set  
is required with OPAL based isochrones to solve this ambiguity.
\par
\begfigps 3.3cm
\psfig{file=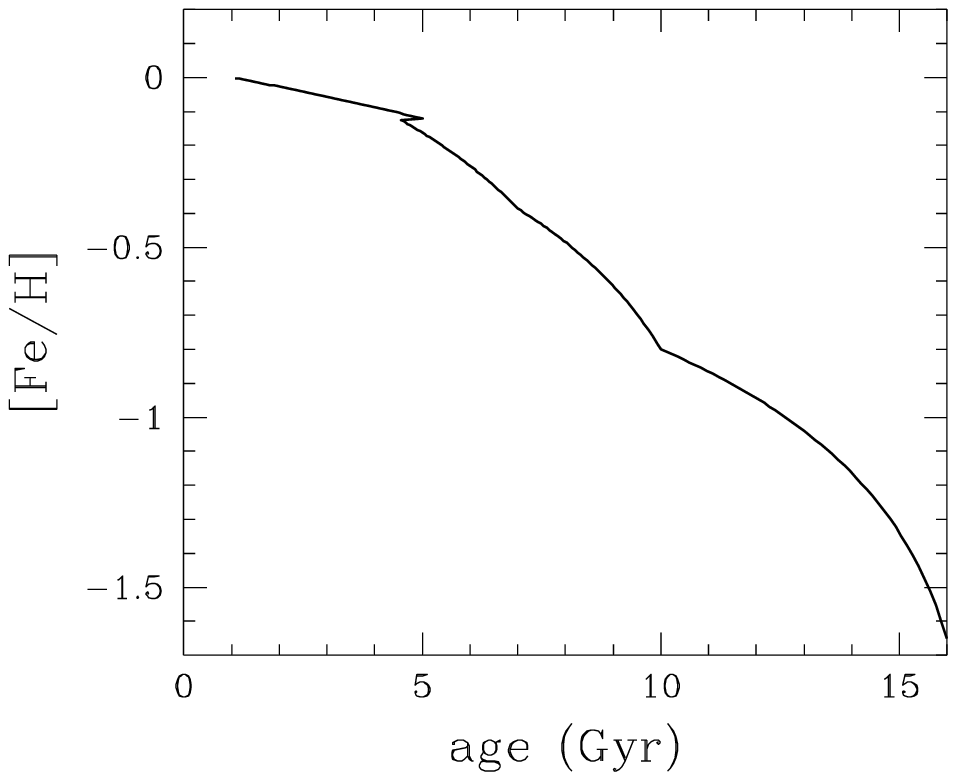,height=5.2cm,width=8.6cm}
\figure{7}{The age-metallicity relation for the disc populations
in Table 1 
}
\endfig
In the radial direction there is apparently no 
metallicity gradient within a population,
which is in agreement with  
the suggestions obtained from B-stars
(Kaufer et al. 1994, Prantzos \& Aubert 1995).
Due to their short lifetimes, B-stars can be considered as 
one population.
On the other hand, Carraro \& Chiosi (1994) and Piatti et al. (1996)
showed with young open clusters that there is a 
radial metallicity gradient.
\par\noindent
If one samples from a young to an older population
and if the scale length is the same for all populations
one would find no metallicity gradient. 
On the other hand,  
an inter-population metallicity gradient should be present
when the various disc populations
do not have the same scale length.
The direction of this gradient is directly
related with an increasing 
or a decreasing disc scale length towards 
younger stellar populations. 
The metallicity gradient from planetary nebulae
(Maciel \& Koppen 1994 and references cited therein)
might has its origin in a mix of the disc populations
and it could be an indication for a different scale length
of these populations.
An independent indication that this is the case 
comes from open clusters (Paper III), which gives 
$h_0=2.3+0.12\,t_9$~kpc ($t_9$ the age of the population
in Gyr).
The scale length determined for the young \& thick disc component,
respectively \hbox{2.5\muspc$\pm$\muspc0.3}~kpc (Robin et al. 1992)
\& \hbox{2.8\muspc$\pm$\muspc0.8}~kpc (Robin et al. 1996) 
or \hbox{2.3\muspc$\pm$\muspc0.6}~kpc \& \hbox{3.8\muspc$\pm$\muspc0.5}~kpc
(Ojha et al. 1996), provide another indication.
The 2.8~kpc scale length from Robin et al. (1996) is probably too small.
It is based only on the field towards the galactic centre,
while their parameter space towards the GAC
field appears to be degenerate (see the likelihood contours in
their Figs.~5a\to{d}). 
Our results favour the presence 
of a radial metallicity gradient induced by the subtle difference
of the scale length of different stellar populations.
\par
Perpendicular to the galactic plane the following situation occurs:
within a particular disc population there is no metallicity gradient, 
while sampling from young (relatively nearby) 
to older (relatively far away) populations will give
rise to a vertical metallicity gradient.
This gradient might be stronger than the radial gradient and is in fact 
in agreement with the results found by Piatti et al. (1996).
These results do not contradict the result obtained by Carraro \& Chiosi 
(1994). They found no evidence for 
a vertical gradient after correction of the radial gradient,
i.e. they found no gradient within a population. 
\par
\hyphenation{Stobie}
\begfigps 3.3cm
\psfig{file=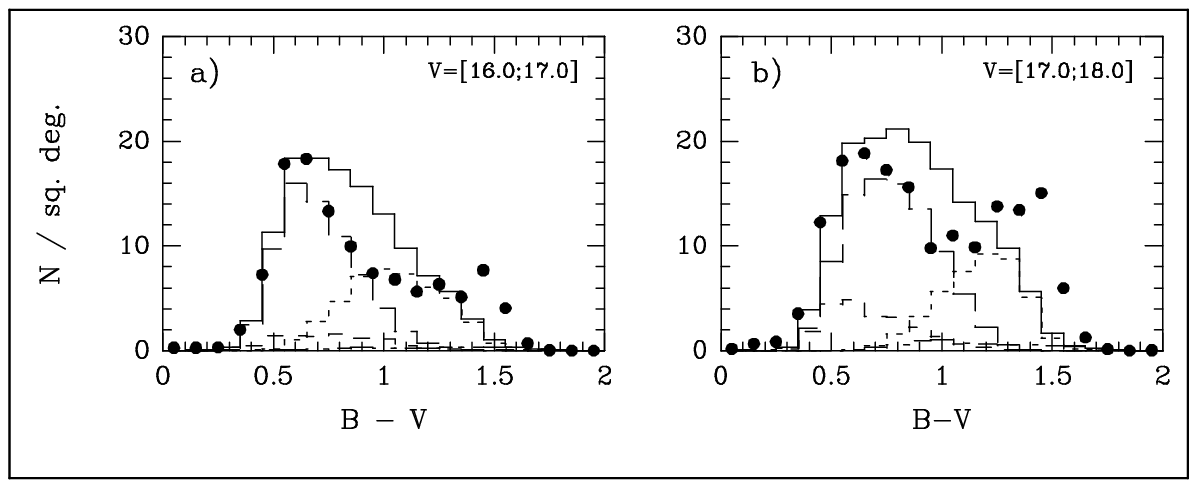,height=3.3cm,width=8.6cm}
\figure{8}{Colour distribution for stars with 
V\muspc=\muspc16\mag0\to18\mag0
from Stobie \& Ishida (1987: filled dots)
in the direction of the NGP,
see Fig.~1.1 for additional details}
\endfig
The different ages of the stellar populations in the disc
are not merely an empirical match to the observations.
It has been observed in the distribution of open clusters 
in the galactic plane. The $\delta$V distribution of open clusters
from Phelps
et al. (1994) shows an almost flat distribution
with two distinct peaks at the end of each side.
The index $\delta$V is an age parameter and is defined as 
the magnitude difference between the MS turn-off and the RHB.
Probably the peaks are related with the occurrence of two burst 
during the formation of the disc. The ages of these clusters 
(Carraro \& Chiosi 1994; Phelps et al. 1994) 
show that these peaks correspond with
burst between 7\to6 Gyr and 2\to1 Gyr. The peak between 
7\to6 Gyr corresponds with the start of the 
formation of the intermediate disc. The peak between 
2\to1 Gyr is related to the formation 
of the young disc population. 
\par

\titleb{Initial mass function}
The colour distributions from RD90 (Figs. 1.2d \& 1.2e),
RM93 (Figs.~1.3a \& 1.3b), SI87 (Figs.~8a \& 8b) 
and GRH85 (Figs.~4.1e \& 4.1f) show
a systematic discrepancy between
the observed and the modeled (\hbox{B--V\muspc$\ga$\muspc0\mag7}) 
colour distributions.
The origin of this difference resides inside the HRD-GST,
because the observations are from independent sources.
From Figs. 2.1\to2.3 it can be seen that 
this is due
to the MS from the disc stars.
Apparently there are too many MS stars generated with 
\hbox{M/M$_\odot\!\!<$\muspc1.0}. 
This contributes partly to the discrepancy
noted between the observed and the simulated colour distributions
in Paper I.
We adopted a slope \hbox{$\alpha$\muspc=\muspc2.35}
for the power-law IMF. The results from Kroupa et al. (1993)
show that this is a reasonable value for the mass range considered here 
(\hbox{M/M$_\odot$\muspc$>$\muspc0.6} to the main sequence turn-off mass
at a particular age).
The present work indicates that the slope of a power-law IMF 
near the HRD-GST mass limit is flatter than 
the slope near the main sequence turn-off.
This is in agreement
with the decrease in the IMF slope 
expected on theoretical grounds (Richtler 1994
and references cited therein). 
\par

\titleb{The solar offset}
Since the early Sixties, the asymmetry in the distribution 
between the NGP and SGP has been ascribed to the solar
position in the plane. The general trend is that an offset,
ranging from 5 -- 40 pc, 
north of the galactic plane has been found (Cohen 1995; 
Humphreys \& Larsen 1995; Paper I).
\par
There would be no difference between the NGP \& SGP data set,
if the Sun is located exactly in the galactic midplane.
A comparison of the model predictions
and the observations between these fields show some differences,
which is likely due to the solar offset.
Discrepancies between the NGP \& SGP is then minimized
by putting the Sun with 10~pc steps out of the galactic plane.
The SGP \& NGP data sets from respectively
GRH85 and SI87 have been used,
because they have comparable photometric errors. 
An integral approach was made, but
emphasis is put on 
the bright stars \hbox{(V\muspc$<$\muspc14\mag0)} in these fields.
With this procedure a determination was made of 
the displacement of the Sun from the galactic plane
and an offset 
\hbox{45\muspc$\pm$\muspc5~pc} north from the galactic plane 
is obtained. This is higher than the generally accepted value 
around 15~pc
(Conti \& Vacca 1990,
\hbox{Cohen~1995}, Hammersley et al. 1995, Binney et al. 1996). 
On the other hand, it is remarkably similar
to the values obtained by SI87 and YY92.
The data sets used for this determination are in fact the same,
but different methods are used.
The large value is probably due 
to systematic differences in an heterogeneous data sets. 
For example, small differences in the zero point between 
the photometric systems or their transformations to the
standard Johnson \hbox{B~\&~V} magnitudes might be responsible for the 
high offset. 
For a better determination of the offset a
homogeneous data set ought to be preferred.
In a photographic survey
Humphreys \& Larsen (1995) find from such a data
set a lower value for the solar offset. With a scale height
of 350~pc they obtained 
\hbox{20.5\muspc$\pm$\muspc3.5~pc}. 
Humphreys \& Larsen (1995) over-estimated the 
scale height of the major disc component, which is
250 pc (Kent et al. 1991, Ng 1994, Papers I \& III).
In that case, they would have obtained 
with their Fig.~6 a value
of about 15 pc. 
\par
The different methods all converge to an offset  
of about 15~pc north from the galactic plane when a homogeneous
data set is used.
It is therefore even more striking that 
\hbox{13.5\muspc$\pm$\muspc1.9~pc} was 
found by van Tulder (1942) from detailed studies of 
the distributions of various types of stars.
It is concluded
that the solar offset is around 15~pc.
It is stressed that this result
does not influence any
of the other results presented in this paper,
because this effect is small on the stars from
the old and thick disc population. 
This offset is important though in the studies of
(very) young stars in the galactic plane. 
\par

\titleb{Stellar populations in the disc}
In Paper~I it is found, that at low latitudes
the dominating populations must be of intermediate 
to young age. For these populations a scale height 
of respectively 250~pc and 100~pc is found.
From the analysis of the NGP data these results are
confirmed. 
The population with a scale height of 500~pc in our analysis 
is called the old disc.
At high galactic latitudes a fourth metal-poor
population with a 1.0~kpc scale height is found.
This population is hereafter referred to as the thick disc.
\par

\titlec{The old disc population}
From the analysis of $ubvy\beta$ CCD photometry 
of stars brighter than V\muspc=\muspc20$^m$ in a field
($l\!=\!262^\circ,b\!=\!4^\circ$)
close to the galactic plane J{\o}nch-S{\o}rensen
\& Knude (1994) found a separate population of stars.
At a distance of about 10 kpc, about 700 pc above the plane,
a significant number of these stars are found. 
The age and chemical composition of these stars are well 
represented by an 8~Gyr isochrone with [Fe/H]\muspc=\muspc\to0.47
(Z\muspc$\simeq$\muspc$0.005$). According to J{\o}nch-S{\o}rensen
\& Knude (1994) these stars are part of the so-called extended disc
(Majewski 1993). 
The contribution from stars 
with a scale height of 1.4\to1.6 kpc (RM93)
in the extended/thick disc population
is expected to be the dominant population, starting 
at about 20 kpc distance
along the line of sight of this field.
The stars in question though are located at half this distance. 
The age and metallicity 
are furthermore in good agreement with,
what is called in this paper, the old disc population,
which has a scale height of 500 pc.
Therefore, the population identified by J{\o}nch-S{\o}rensen
\& Knude (1994) is strongly related with our old disc population. 
The results from Paper~III ($l\!=\!1\fdg0,b\!=\!-3\fdg9$)
%from the analysis of the CMDs from Baade's Window
%(Paczy\'nski et al. 1994) with the HRD-GST 
indicate that a small
feature near 
V--I\muspc$\simeq$\muspc1\mag2\to1\mag4 and 
V\muspc=\muspc19$^m$ in the CMDs is related with the MS turn-off 
of this disc population. The consistency between fields in different
directions in the galactic plane are an indication that the 
old disc population is distinct in age-metallicity and 
scale height.
\par

\titlec{The age-metallicity of the thick disc}
In Figs. 1.2e, 1.2f, 1.3b, and 1.3c it is noted 
that the colour distributions from the MC simulations
do not match the observed distributions 
at B--V\muspc$<$\muspc0\mag3.
Because this trend is present in two independent data sets
it must be a real feature, which is not covered properly
by the HRD-GST. 
Part of the shift in Figs. 1.2e \& 1.2f is due to the gradient 
in the magnitude scale (Sect. 3.1). 
This trend is visible at V\muspc$>$\muspc18$^m$
and is due to the halo or
the thick disc (see Figs. 2.1\to2.6). 
\hfill\break 
The evolutionary tracks in the HRD-GST do not extend to
metallicities lower than \hbox{Z\muspc=\muspc0.0004}. 
A metallicity \hbox{Z\muspc$<$\muspc0.0004} 
will not account fully for the discrepancy, because
this will only shift a small number of HB (horizontal 
branch) stars to bluer colours.
One should be aware, that due to the metallicity range 
and the implicit linear age-metallicity relation in the 
HRD-GST (Paper~I), the majority of
the stars have metallicities larger than \hbox{Z\muspc=\muspc0.0004}.
Therefore, the origin for this discrepancy must be due to the age 
of one or both populations. If the halo population
is younger, the simulated distributions 
at fainter magnitudes (frames d, e, and~f from Fig.~1.3)
becomes bluer. The observed distributions 
do not allow this. Therefore, the thick disc population 
is likely younger than assumed. An upper limit for the age in the interval
13\to16 Gyr is well possible.
Such an age gives a small contribution of `blue' stars between 
16$^m\!<\!V\!<\!18^m$.
It is not clear though if this 
removes completely the discrepancy between 
18$^m\!<\!V\!<\!19^m$.
\par
The metallicity of the thick disc is a controversial issue. 
Morrison et al. (1990) found from metallicity ranking 
with DDO photometry evidence for a metal-poor tail. 
The spectroscopic results from Ryan \& Lambert (1995) 
on a sample of stars, classified as metal-poor with DDO photometry, 
contradict this, but they lack the data to confidently 
reject the presence of metal-poor stars in a thick disc.
Their sample is formed by giant stars with V magnitudes spanning the
range 8\mag0\to12\mag0 and B--V colours ranging roughly 
from 0\mag75\to0\mag95.
Figures 2.2\to2.4 show that with these criteria virtually no 
metal-poor stars from the thick disc would have entered their sample. 
Their result mainly demonstrate 
that metallicity ranking with DDO photometry is not reliable 
towards low metallicities.
\par\noindent
Beers \& Sommer-Larsen (1995) performed a larger spectroscopic study 
and found convincing evidence for a metal-poor tail for the thick disc 
down to at least \hbox{[Fe/H]\muspc=\muspc--2.0}.
They selected a large variety of stars (MS, RGB, HB 
and AGB stars), covering a large magnitude interval. 
Their study does not contain the selection bias as in the study
from Ryan \& Lambert. In conclusion: our findings for
a metal-poor thick disc are in agreement with  
the results from Beers \& Sommer-Larsen.

\titlec{The thick disc population: scale height}
The metal-poor thick disc is identified from the excess of stars 
at the blue edge (B--V\muspc$\simeq$\muspc0\mag4) 
of the colour distributions in Figs.~1.2d, 1.2e, 1.3a \& 1.3b. 
The parameters found here are different from those 
obtained with the Besan\c con model 
(Robin et al. 1996, Ojha et al. 1996).
Their thick disc parameters  
are \hbox{$z_0$\muspc=\muspc760\muspc$\pm$\muspc50 pc}
and [Fe/H] ranging from --\muspc0.6 to --\muspc0.9 
or \hbox{Z\muspc$\simeq$\muspc0.003\to0.005}. These values 
are intermediate 
to the parameters determined for our old and thick disc population.
%Ojha et al. (1996) selected the stars in the height range
%\hbox{150\muspc$<$\muspc$z$\muspc$<$\muspc2750 pc}.
Their metallicity is close to our transition metallicity
between the two populations. 
From our work we expect, that the age of their thick disc 
population is close to 10~Gyr.
\par
The difference in the results originates from the initial conditions
between the two models. The Besan{\c c}on model adopts a vertical
age-velocity dispersion. In suitably chosen 
age-velocity bins the scale heights are calculated from the 
galactic potential (Bienaym\'e et al. 1987).
These disc sub-populations are then fitted to the observations
and the thick disc is obtained from the stars not fitted by these 
sub-populations. 
We explored the parameter space through 
a direct comparison between the observed and the synthetic distribution.
The age-metallicities and scale heights of the populations described 
in this paper are not expected to coincide exactly with the
selection of age-metallicity bins selected for the Besan{\c c}on model. 
Our results are therefore complementary with those obtained with 
the Besan{\c c}on model.
This explains why their thick disc scale height can be intermediate to our 
old and thick disc value, but this does not explain the actual value.
\par
About half of the star counts surveys used by Robin et al.
(1996) are complete up to about \hbox{V\muspc=\muspc18$^m$}.
At this magnitude a disc population will have a scale height intermediate
to our old and thick disc (see Figs. 2.3 \& 2.4), 
i.e. \hbox{$z_0$\muspc$\simeq$\muspc750~pc}.
For a small number of data sets the limiting magnitude was brighter
than \hbox{V\muspc=\muspc18$^m$} and these data sets cannot
be used to constrain a disc population with 
\hbox{$z_0$\muspc$\ga$\muspc750~pc}.
Only the RM93 data set is reliable to magnitude limits
fainter than \hbox{V\muspc=\muspc18$^m$},
but the results obtained by Robin et al. 
from this field could be biased by the results obtained
from the other fields.
The scale height for the thick disc 
obtained by Robin et al. (1996) is likely induced by the
completeness limits of the data sets analyzed by them. 
\par
The parameters for the thick disc population listed in Table~1
are not very different from those given by YY92.
They obtained 900~pc, which is within the estimated 
uncertainty of our population.
It is not clear though if the scale height
is comparable with the 1170~pc thick disc 
parameterized by Chen (1996),
but the scale height is within the uncertainties of 
the \hbox{1140\muspc$\pm$\muspc60~pc}
obtained by Spagna et al. (1996).

\titlec{Indications for a possible merger}
The simulations from
Mihos \& Hernquist (1996) indicated that `major' mergers, between
galaxies with comparable mass, should produce objects 
resembling ellipticals. Quinn et al. (1993) show that
a merger event with galaxies comparable to the LMC and SMC could destroy 
the parent disc. 
%A thick disc would form 
%with a scale height, which is roughly two times larger 
%than its `parent' population. 
%Star formation in a post-merger disc results in a younger population
%of stars, with a scale height smaller than the thick disc. 
%If one considers only age-metallicity and scale height, this result
%is indistinguishable from a hierarchical formation process. 
%A merger event tends to modify the features to be encountered
%from a hierarchical formation process. 
%In this respect,
%the kinematics of the stars are essential to distinguish
%the two scenarios from each other.
Unavane et al. (1996) demonstrated quantitatively, that 
a merger scenario is not likely between our Galaxy and 
a LMC or SMC type galaxy.
If a merger event had occurred, it should have been with a 
galaxy with a mass comparable to those from dwarf ellipticals. 
Walker et al. (1996) show that `minor' merger events with 
dwarf elliptical satellites do not destroy the parent disc and 
that they could puff up part of the disc by \hbox{$\sim$\muspc60\%}.
It is not clear, if the puffed up part of the disc forms a layer on top
of the parent disc. And if this layer 
is in age-metallicity and scale height distinct from a hierarchical 
formation process.
Walker et al. argue that other arrangements are 
certainly possible with the parameters
available to twiddle the simulations.
Their work provide indications about general aspects 
of merger events. 
Detailed numerical inference from these simulations should be avoided,
because the results depend sensitively on the choice of scale parameters.
\par
A minor merger event could move some stars from the
metal-poor disc population with a 1~kpc scale height to roughly
a 2~kpc scale height. The RR-Lyrae stars
from a flattened component with a 2~kpc scale height
from Kinman et al. (1994)
could be a  record from a fossil merger event. 
The RM93 thick disc is possibly another record of a minor 
merger event. The 1.2\to1.4~kpc scale height is larger than the 1.0~kpc 
scale height for the metal-poor thick disc component discussed 
in Sects. 4.4.2 \& 4.4.3. On the other hand,
the colours of the stars from the RM93 thick disc are redder. 
It is important to determine how the RM93 thick disc stars are
related in age and metallicity with a population intermediate
to the populations with a 0.5~kpc and 1.0~kpc scale height from our work.
This provides an indication on when a minor merger event 
could have occurred with our Galaxy.
\par

\titlec{Low mass stars}
The effect of the 0.6 M/M$_\odot$ mass limit 
(stars with smaller masses are hereafter referred to as low mass
stars) in the HRD-GST is that 
the features with \hbox{B--V\muspc$>$\muspc0\mag95} 
in Fig. 1.3f to
\hbox{B--V\muspc$>$\muspc1\mag35}
in Figs. 1.1f \& 1.3a are not covered by the MC simulations.
However, the synthetic CMDs give indications on the contribution of the low
mass stars. 
Figs.~2.1\to2.4 indicate that the small number of missing stars red stars 
in Figs.~1.1f \& 1.3a are related to the young disc population 
with a 100~pc scale height, while the missing stars
in Figs. 1.3b \& 1.3c are due to the combined contribution
of low mass stars from populations with a scale height 
of 100~pc, 250~pc and 500~pc.
The low mass stars from 
populations with a 500~pc and 1.0~kpc scale height 
are the main contributors 
at V\muspc$>$\muspc20$^m$ and \hbox{B--V\muspc$>$\muspc0\mag95}.
The cumulative distribution of the low mass stars can be used 
to constrain the IMF down to the hydrogen burning main sequence
mass limit in a similar way as discussed in Sect.~4.2 for stars
with M/M$_\odot\!>\!0.6$. 
Only in this way, mixes of stellar populations with different 
scale heights, ages and metallicities are taken properly into account. 
According to Mera et al. (1996) the slope 
\hbox{$\alpha$\muspc=\muspc2.0\muspc$\pm$0.5} 
for the power-law IMF for stars with
\hbox{M/M$_\odot\!<$\muspc0.5}, while Kroupa et al. (1993) 
and Tinney (1995) find a considerable smaller value.
\par 

\titleb{Stellar populations in the spheroid}
In the spheroidal component two populations are distinguished.
The first is easily identified with the metal-poor halo population.
The age range of the halo stars is comparable with the 
\hbox{5\to7~Gyr} age range from the bulk of
halo globular clusters (Chaboyer et al. 1996).
The second is a metal-richer bulge population.
Its presence is not established unambiguously from this 
study. The parameterization adopted here is taken from Paper~II,
which differs from Paper~III, mainly in the upper metallicity limit. 
It is not clear though whether the difference between Paper~II \& III 
is caused by an outward, radially decreasing metallicity gradient.
Not enough fields have been analyzed thus far, but the work 
from Minniti (1995 and references cited therein) appears to support this.
Both populations reside in the spheroid, which is 
barely or not flattened. 
\par

\titlec{Flattening of the spheroid}
The hitherto supposed flattening 
of the spheroid is likely due to mixtures of
the metal-poor halo with the thick disc stars. 
The flattening of the spheroid is 
very small ($0.95\!\!<\!\!q\!\!<\!\!1.00$).
A negligible amount of flattening has already been noted in the 
distribution of halo globular clusters (Zinn 1985).
The colour distribution is not matched properly 
with a power-law at \hbox{V\muspc$>$\muspc20$^m$}
for a stronger flattening, which would disperse the halo
stars over a larger magnitude range. This is no in agreement with
the more concentrated distribution of the halo stars.
\par
A duality in the distribution of metal poor stars has been suggested 
by Hartwick (1987) from the analysis of the density distribution
of RR Lyrae stars. 
Similar results are obtained by Norris (1994) and Kinman et al. (1994).
An extensive discussion of this duality 
is given by Norris (1994 and references cited therein). 
He finds in high proper motion samples of MS stars
two distinct components:
the halo and the so-called metal-poor thick disc
(Morrison et al. 1990) 
with 
\hbox{\to3.0\muspc$\le$\muspc[Fe/H]\muspc$\le$\muspc\to1.0}.
These values are
consistent with the metallicity range 
\hbox{Z\muspc=\muspc0.0004\to0.003}
obtained from the analysis 
of the NGP star counts with the HRD-GST (taking into account 
that 
in the current setup of the HRD-GST 
the lower metallicity limit 
is Z\muspc=\muspc0.0004  
and that the upper limit 
can be in the range Z\muspc$\simeq$\muspc0.001\to0.005).
A consequence of this duality is that
RR~Lyrae stars are expected in both the halo and the 
thick disc. A sample of supposed halo \hbox{RR~Lyrae} stars can be
polluted with stars from the thick disc.
Therefore, Wesselink (1987) might have
overestimated the amount of flattening 
from his sample of RR~Lyrae stars, presumably located in the halo.
\par
\begfigps 3.3cm
\psfig{file=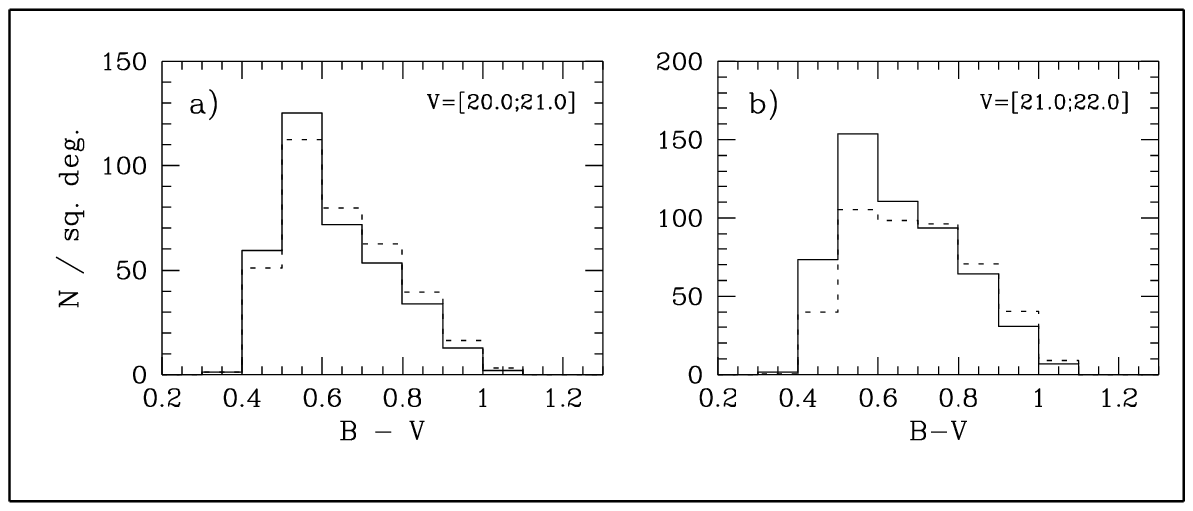,height=3.3cm,width=8.6cm}
\figure{9}{Colour distribution for synthetic
stars from the halo
population in the magnitude interval \hbox{V\muspc=\muspc20\mag0\to22\mag0},
described by a power-law (solid line) and a
\hbox{$R^{1/4}$--law} (dashed line)
in the direction of the NGP}
\endfig
Our results support the notion that an overlap
is present in the age and chemical
composition of the halo and thick disc populations.
They indicate that the classical flattening parameter
in star counts models needs a revision.
Star counts studies towards other deep fields towards the galactic poles 
are required to verify this. 
\par

\titlec{Power-law or R$^{1/4}$--\muspc{law}}
In contrast to a power-law with
index \hbox{$n\!=\!3.0$}, 
the MC simulations show that
the \hbox{$R^{1/4}$--law}, as parameterized by 
Bahcall \& Soneira (1984), does not give for V\muspc$>\!\!20^m$
the proper distribution. In particular for the stars with
B--V\muspc$\simeq$\muspc0\mag5, where
the density starts to drop significantly with a \hbox{$R^{1/4}$--law}.
In Figs. 9a \& 9b the two distributions are compared with
each other. They show that a halo population
described with the \hbox{$R^{1/4}$--law} predicts 
at fainter magnitudes less stars with respect to a power-law description.
This drop in density cannot be fully compensated by adopting 
a stronger flattening, because the drop in the \hbox{$R^{1/4}$--law}
is mainly driven by the exponential in this expression.
Although the number of stars are 
considerably smaller, a similar discrepancy is visible in 
a comparison between the star counts (I\muspc=\muspc23$^m$\to25$^m$)
and a galactic model that utilizes the 
\hbox{$R^{1/4}$--law} for the description
of the halo (Mould, 1996).
Differences are notably also
present in the distribution of the RGB (red giant branch)  stars with 
\hbox{B--V\muspc$>$\muspc1\mag0}.
\par

\titlec{Metal-rich stars}
The contribution of the halo becomes noticeable
at \hbox{V$\,>\,$18$^m$} (see figure 1.3c and 2.6). 
At \hbox{V\muspc$>$\muspc21$^m$} the width of the 
\hbox{B--V} colour distribution
between 0\mag4 to 1\mag0 is too wide  
for halo stars only. 
There is a contribution missing between 
\hbox{0\mag8\muspc$\la$\muspc{B--V}\muspc$\la$\muspc1\mag0}.
The contribution from the old and thick disc population
(see Figs.~2.3 \& 2.4) is not sufficient to explain this.
This missing stars must be fairly old, because any population which
is significant 
younger becomes not only bluer, but also gives
a larger contribution at magnitudes brighter than 
\hbox{V\muspc=\muspc20$^m$}.
An additional contribution could be
due to metal-rich stars from an old population.
The numerical calculations from 
Steinmetz \& M\"uller (1995) suggest the presence 
of some metal-rich stars in the outer parts of the Galaxy. 
In the simulations a population similar to  
the metal-rich population found in Paper~II 
is adopted. 
Figure 1.3f shows, that the inclusion of this population gives
a better agreement between the MC simulations and the observations. 
The presence of metal-rich stars cannot be unambiguously
established, but on the other hand it cannot be excluded.
This shows that the metal-rich stars are not dominantly present in 
the outer parts of the Galaxy.

\titlea{Conclusions \& future work}
We studied the colour
distribution towards the galactic poles and anti-centre.
The analysis shows that the distributions are well reproduced
by the sum of four stellar populations in the disc: 
a young, intermediate, old, and thick disc population.
\par
In the spheroid two major populations are distinguished:
a metal poor halo population and a metal richer `bulge' population.
The presence of the latter could not be well 
constrained from the north galactic pole observations.
Both the halo and the bulge are well described by a power-law with index 
\hbox{$n$\muspc=\muspc3.0}. Furthermore, the spheroid appears not
to be flattened or the flattening is very small 
\hbox{(0.95\muspc$<\!\!q\!\!<$\muspc1.00)}.
\par 
There are still some discrepancies present between the 
model and the observed colour distributions. These 
discrepancies are related with the decrease of the slope of 
the power-law IMF for stars with M/M$_\odot\!<$\muspc1.0
and the age of the thick disc stars. 
They might bias the local normalization. Therefore,
the IMF will be the subject for future research, 
where the lower mass limit in the HRD-GST library of stellar 
evolutionary tracks is extended downwards to 0.15~M$_\odot$.
With an improved galactic model the HRD-GST might
be able to obtain better constraints for the bulge population
and to distinguish between stellar populations from a 
hierarchical formation process or those from a merger
event.

\acknow{Dr. A. Blaauw is thanked for his helpful \& critical notes 
on Chapter 8 of Ng's Ph.D. thesis. 
This is a revised and updated version of this chapter.
Ng acknowledges 
J. Lub and D.K. Ojha for helpful discussions \& comments and 
M. Cohen, A. Koekemoer, N. Reid and A.C. Robin for 
their suggestions.
G. Bertelli, A. Bressan and C. Chiosi acknowledge
the financial support received from the Italian Ministry of University,
Scientific Research and Technology (MURST) and the 
Italian Space Agency (ASI).
Ng acknowledges the financial support, received at the IAP-CNRS,
from HCM grant CHRX-CT94-0627 from the European Community.}

\begref{References}
%\ref Arp H., 1965, ApJ 141, 43
\ref Bahcall J.N., Soneira, R.M., 1984, ApJS 55, 67
\ref Beers T.C., Sommer-Larsen J., 1995, ApJS 96, 175
\ref Bertelli G., Mateo M., Chiosi C., Bressan A., 1992, ApJ 388, 400
\ref Bertelli G., Bressan A., Chiosi C., Fagotto F., Nasi E.,
1994, A\&AS 106, 275
\ref Bertelli G., Bressan A., Chiosi C., Ng Y.K., Ortolani S.,
    1995, A\&A 301, 381 (Paper II)
\ref Bertelli G., Bressan A., Chiosi C., Ng Y.K., 1996, A\&A 
     310, 115
\ref Bienaym\'e O., Robin A.C., Cr\'ez\'e M., 1987, A\&A 180, 94
\ref Binney J., Gerhard O., Spergel D., 1996, MNRAS {\it submitted}\/
({\tt astro-ph/9609066})
\ref Blaauw A., Elvius T., 1965, Stars and Stellar Systems Vol. 5,
       ``Galactic Structure'', A. Blaauw and M. Schmidt eds., 589 
%\ref Blommaert J.A.D.L., van der Veen W.E.C.J., Habing H.J., 1993, 
%     A\&A 267, 93
\ref Carraro G. Chiosi C., 1994, A\&A 287, 761
\ref Chaboyer B, Demarque P., Sarajedini A., 1996, ApJ 459, 558
\ref Chen B., 1996, A\&A 306, 733
\ref Chiu L.T.G., 1980, ApJS 44, 31
\ref Cohen M., 1995, ApJ 444, 874
\ref Conti P.S., Vacca W.D., 1990, AJ 100, 431
\ref Edvardsson B., Andersen J., Gustaffson B., et al., 1993,
A\&A 275, 101
%\ref Eggen O.J., Lynden-Bell D, Sandage A.R., 1962,
%     ApJ 136, 748
\ref Faber S.M., Burstein D., Tinsley B., King I.R., 1976,
     AJ 81, 45 
\ref Fenkart R., Esin-Yilmaz F., 1983, A\&AS 54, 423 
\ref Freeman K.C., 1992, IAU Symposium 149, in `The stellar populations
of galaxies', B. Barbuy and A. Renzini (eds.), 65
\ref Friel E.D., Cudworth K.M., 1986, AJ 91, 293
%\ref Fux R., Martinet L., 1994, A\&A 287, L21
\ref Gilmore G., Reid N., Hewett P., 1985, MNRAS 213, 257 (GRH85)
\ref Girardi L., Bressan A., Chiosi C., 1996a, Proceedings 
     `Stellar evolution: what should be done', 3\to5 July 1995,
     Li\`ege (Belgium), A Noels et al. (eds.), 39
\ref Girardi L., Bressan A., Chiosi C., Bertelli G., Nasi E.,
     1996b, A\&AS 117, 113
\ref Hammersley P.L., Garz\'on F., Mahoney T., Calbet X., 1995,
MNRAS 273, 206
\ref Hartwick F.D.A., 1987, in `The Galaxy', G. Gilmore and
B. Carswell (eds.), Dordrecht: Reidel, 281
%\ref van den Hoek L.B., de Jong T., 1996, A\&A {\it accepted}
\ref Huebner W.F., Merts A.L., Magee N.H., Argo M.F., 1977,
Los Alamos Scientific Laboratory Report LA-6760-M
\ref Humphreys R.M., Larsen J.A., 1995, AJ 110, 2183
\ref Iglesias C.A., Rodgers F.J., Wilson B.G., 1992, ApJ 397, 717
\ref J{\o}nch-S{\o}rensen H., Knude J., 1994, A\&A 288, 139
\ref Kaufer A. Szeifert Th., Krenzin R., Baschek B.,
Wolf B., 1994, A\&A 289, 740
\ref Kent S.M., Dame T., Fazio G., 1991, ApJ 378, 131
\ref Kinman T.D., Suntzeff N.B., Kraft R.P., 1994, AJ 108, 1723 
\ref Koo D.C., Kron R.G., 1982, A\&A 105, 107
\ref Kron R.G., 1980, ApJS 43, 305
\ref Kroupa P., Tout C., Gilmore G., 1993, MNRAS 262, 545
%\ref Lee Y.W., 1992, Proceedings IAU symposium 149, 
%`Stellar Populations of Galaxies', 4\to8 August 1991, Angra dos Reis
%(Brazil), B. Barbuy and A. Renzini (eds.), 446
\ref Maciel W., Koppen J., 1994, A\&A 282, 436
\ref Majewski S.R, 1992, ApJS 78, 87
\ref Majewski S.R, 1993, ARA\&A 31, 575
%\ref Majewski S.R., Munn J.A., Hawley S.L., 1994, ApJ 427, L37
\ref Mera D., Chabrier G., Baraffe I., 1996, ApJ 459, L87
\ref Minniti D., 1995, AJ 109, 166
\ref Mihos J.C., Hernquist L., 1996, ApJ 464, 641
\ref Morrison H.L., Flynn C., Freeman K.C., 1990, AJ 100, 1191 
\ref Mould J., 1996, PASP 108, 35
\ref Murray C.A., Argyle R.W., Corben P.M., 1986, MNRAS 223, 629
\ref Ng Y.K., 1994, Ph.D. thesis, Leiden University, the Netherlands
\ref Ng Y.K., 1996, Proceedings
     `The impact of large-scale near-IR sky surveys',
     24\to26 April 1996, Puerto de la Cruz (Tenerife; Spain),
     P. Garzon-Lopez (ed.), {\it in press}
\ref Ng Y.K., Bertelli G., Bressan A., Chiosi C., Lub J.,
    1995, A\&A 295, 655 (Paper I; erratum A\&A 301, 318)
\ref Ng Y.K., Bertelli G., Chiosi C., Bressan A., 1996a, A\&A 
     310, 771 (Paper III)
\ref Ng Y.K., Bertelli G., Chiosi C., Bressan A., 1996b,
     `Spiral Galaxies in the near IR',
     Garching bei M\"unchen, 7--9 June 1995, D. Minniti 
     and H.-W. Rix (eds.), 110
\ref Norris J.E., 1994, ApJ 431, 645
\ref Ojha D.K., Bienaym\'e O., Robin A.C., Mohan V., 1994, 
     A\&A 284, 810 (OBRM94)
\ref Ojha D.K., Bienaym\'e O., Robin A.C., Cr\'ez\'e M., Mohan V., 1996,
     A\&A 311, 456
%\ref Ortolani S., Renzini A., Gilmozzi R., Marconi G., Barbuy B.,
%     Bica E., Rich R.M., 1995, Nature 377, 701
\ref Paczy\'nski B., Stanek K.Z., Udalski A., et al., 
%Szyma\'nski, M., Ka{\l}u{\zdot}ny, J., Kubiak, M., Mateo, M., 
1994, AJ 107, 2060
%\ref Perry C.L., 1969, AJ 74, 139
%\ref Perryman M.A.C., Lindegren L., Kovalensky J., 1995, A\&A 304, 69
\ref Phelps R.L., Janes K.A., Montgomery K.A., 1994, AJ 107, 1079
\ref Piatti A.E., Claria J.J., Abadi M.G., 1995, AJ 110, 2813
\ref Prantzos N., Aubert O., 1995, A\&A 302, 69
\ref Rodgers F.J., Iglesias C.A., 1992, ApJS 79, 507
\ref Quinn P.J., Hernquist L., Fullager D.P., 1993, ApJ 403, 74
\ref Reid N., 1990, MNRAS 247, 70 (RD90)
\ref Reid N., Majewski S.R., 1993, ApJ 409, 635 (RM93)
%\ref Renzini A., 1993, Proceedings IAU symposium 153, 
%`Galactic Bulges', 17\to21 August 1992, Ghent
%(Belgium), H. Dejonghe and H. Habing (eds.), 325
\ref Richtler T., 1994, A\&A 287, 517
\ref Robin A., Cr\'ez\'e M., 1986, A\&AS 64, 53
\ref Robin A.C., Cr\'ez\'e M., Mohan V., 1992, A\&A 253, 389
\ref Robin A.C., Haywood M., Cr\'ez\'e M., Ojha D.K., Bienaym\'e O.,
1996, A\&A 305, 125
%\ref Rocha-Pinta H.J., Maciel W.J., 1996, MNRAS 279, 447
%\ref Rodgers A.W., 1971, AJ 165, 581
%\ref Rodgers A.W., Roberts W.H., 1993, AJ 106, 1839
\ref Ryan S.G., Lambert D.L., 1995, AJ 109, 2068
%\ref Searle L., Zinn R., 1978, ApJ 225, 357
\ref Spagna A., Lattanzi M.G., Lasker B.M., et al., 
1996, A\&A 311, 758
\ref Stobie R.S., Ishida K., 1987, AJ 93, 624 (SI87)
\ref Soubiran C., 1992, A\&A 259, 394
\ref Steinmetz M., M\"uller E., 1995, MNRAS 276, 549
\ref Tinney C.G., 1995, ApJ 445, 1017
\ref van Tulder, J.J.M., 1942, Bull. Astron. Inst. Netherlands 9, 315
\ref Unavane M., Wyse R.F.G., Gilmore G., 1996, MNRAS 278, 727
\ref Vandenberg D.A., 1985, ApJS 58, 711
\ref Walker I.R., Mihos J.C., Hernquist L., 1996, ApJ 460, 121
\ref Weistrop D., 1972, AJ 77, 366
\ref Wesselink Th.J.H., 1987, Ph.D. thesis, Catholic University
of Nijmegen, the Netherlands
\ref Yamagata T., Yoshii Y., 1992, AJ 103, 117 (YY92)
\ref Yoshii Y., Ishida K., Stobie R.S., 1987, AJ 93, 323
\ref Zinn R., 1985, ApJ 293, 424
\endref

\bye